\documentclass[aps,prb,twocolumn,amsfonts,showpacs,longbibliography,citeautoscript,superscriptaddress]{revtex4-2}
\usepackage[toc,page]{appendix}
\usepackage[british]{babel}
\usepackage[utf8]{inputenc}
\usepackage{amsmath,amssymb}
\usepackage{graphicx}
\usepackage{ae}
\usepackage{esdiff}
\usepackage{braket}
\usepackage{bbm}
\usepackage{simplewick}
\usepackage{float}
\usepackage{braket}
\usepackage{simplewick}
\usepackage{color}
\usepackage{framed}
\usepackage[caption=false]{subfig}
\usepackage[pdftex,colorlinks=true,linkcolor=blue,citecolor=blue,urlcolor=black]{hyperref}
\usepackage{hyperref}
\usepackage{mathtools}
\usepackage{enumitem}
\usepackage{ragged2e}
\usepackage{floatflt}
\usepackage{import}
\usepackage{titlesec}
\usepackage{multirow}
\usepackage{etoolbox}
\usepackage{appendix}
\usepackage{dsfont}
\usepackage{comment}
\usepackage{xcolor}
\usepackage{blindtext}
\usepackage[makeroom]{cancel}
\usepackage{mathtools}

\begin{document}

\newcommand{\norm}[1]{\left\lVert#1\right\rVert}

\def \r{{\boldsymbol{r}}}
\def \k{{\boldsymbol{k}}}
\def \p{{\boldsymbol{p}}}
\def \q{{\boldsymbol{q}}}
\def \x{{\textbf{x}}}
\def \A{{\textbf{{A}}}}
\def \a{{\textbf{{a}}}}
\def \b{{\textbf{{b}}}}
\def \c{{\textbf{{c}}}}
\def \z{{\textbf{{z}}}}
\def \0{{\boldsymbol{{0}}}}
\def \dl{\frac{\partial}{\partial l}}
\def \P{{\boldsymbol{P}}}
\def \K{{\boldsymbol{K}}}
\def \sigmad{{\sigma_{\downarrow\uparrow}}}
\def \uone{{\boldsymbol{u}_1}}
\def \utwo{{\boldsymbol{u}_2}}
\def \edown{{\epsilon_{\downarrow}}}
\def \omfl{{\omega_{\text{FL}}}}
\def \piph{\Pi_\text{ph}}
\def \sign{ \text{sign}}
\def \lamt{\tilde{\lambda}}
\def \Sigmab{\Sigma_{\omega^2, \text{bs}}}
\def \intk{{\int_\textbf{k}}}
\def \Ims{\text{Im} [ \Sigma^R(\omega) ]}
\def \Gammabs{\Gamma_{\text{bs}}}
\def \aq{|\q|}
\def \ak{|\k|}
\def \omt{\omega_{\text{th}}}

\definecolor{mgrey}{RGB}{63,63,63}
\definecolor{mred}{RGB}{235,97,51}
\newcommand{\mg}[1]{{\color{mgrey}{#1}}}
\newcommand{\mr}[1]{{\color{mred}{#1}}}

\newcommand{\beq} {\begin{equation}}
\newcommand{\eeq} {\end{equation}}
\newcommand{\bea} {\begin{eqnarray}}
\newcommand{\eea} {\end{eqnarray}}
\newcommand{\be} {\begin{equation}}
\newcommand{\ee} {\end{equation}}
\newcommand{\red}[1]{{\color{red}{#1}}}

\def\BigColSep{\setlength{\arraycolsep}{50pt}}

\title{
Quasiparticle scattering in a superconductor near a nematic critical point: resonance mode and multiple attractive channels.}

\author{Dimitri Pimenov}\email{dpimenov@umn.edu}
\author{Alex Kamenev}
\author{Andrey V.\ Chubukov}
\affiliation{William I. Fine Theoretical Physics Institute, University of Minnesota, Minneapolis, MN 55455, USA}

\begin{abstract}
We analyze the scattering rate for 2D fermions interacting via soft nematic fluctuations. The ground state is an $s$-wave superconductor, but other pairing channels are almost equally attractive. This strongly alters the scattering rate: At energies beyond the pairing gap $\Delta$, it is renormalized by contributions from all pairing channels. At energies of order $\Delta$, it is determined by the competition between scattering into a gapped continuum and dispersing nematic resonance. The outcome is a ``peak-peak-dip-hump" spectrum, similar, but not identical, to the ``peak-dip-hump" structure in the cuprates.
\end{abstract}
\maketitle

\textbf{Introduction}.---
The spin resonance mode, observed in a number of quasi-2D cuprate superconductors at momentum $(\pi,\pi)$, has attracted much attention in the studies of high-temperature superconductivity~\cite{Fong1996,Dai1999,Bourges2000}.
 The resonance mode mediates a 4-fermion interaction and accounts for
the ``peak-dip-hump'' structure of the fermionic spectral function observed in ARPES measurements~\cite{PhysRevLett.83.3709, Shen1998,PhysRevLett.87.067005,Valla1999,Kaminski2000,Fink2006}.
In an itinerant approach, the resonance can be viewed as a spin exciton, whose emergence below the threshold for the continuum is associated with the $d$-wave symmetry of a gap function (see e.g.,~\cite{
 doi:10.1080/00018730600645636,ABANOV2001129} and references therein).

In this Letter, we argue that a similar phenomenon takes place in a 2D $s$-wave superconductor near a quantum critical point (QCP) towards a charge nematic order. We
demonstrate that
the nematic propagator possesses a charge exciton at a frequency $\Omega_{\text{res}}
(\q)$, which near an instability is far below the threshold for the continuum at twice the gap value, $2\Delta$. A nematic resonance at $\q=0$  has been observed in the $B_{1g}$ Raman spectra of several iron-based superconductors  \cite{PhysRevLett.116.017001,PhysRevB.93.054515, PhysRevB.93.075106, PhysRevB.95.085125,PhysRevB.101.085102};
  here we show that $\Omega_{\text{res}} (\q)$ and   the spectral weight of the resonance  are strongly $\q$-dependent, which is tied to the $(\omega/\aq)$-form of the Landau damping of a  nematic propagator in the normal state. We analyze how the  dispersing  charge exciton affects the fermionic spectral function at energies   above the threshold at $\Delta + \Omega_{\text{res}} (\q=0)$. We find that the spectral function is highly non-monotonic  due to strong momentum dependence of $\Omega_{\text{res}} (\q)$.
In contrast, a  spin resonance in a $d$-wave superconductor does not lead to such behavior as  there
 relevant momenta $\q$  are close to $(\pi, \pi)$.

  We further analyze the form of the spectral function at energies above $3\Delta$, when the key contribution comes from the continuum in the
     nematic propagator.
     Here,
      we reveal additional physics of a  nematic QCP.  Namely,  for an isotropic  fermionic dispersion, the interaction, mediated by nematic fluctuations, yields an attraction in all pairing channels.
  The $s$-wave coupling is the
   largest one, but only by a small number, proportional to the distance to a nematic transition~\cite{PhysRevLett.114.097001, KleinNpj2019}.
    The continuum part of the order parameter propagator  has contributions from all pairing channels, and the fermionic self-energy and the spectral function due to interaction with the continuum are very different from the ones expected for a pure $s$-wave superconductor.
       We single out the $s$-wave contribution and show that it only accounts for a small portion of the full self-energy. We show that
   at even larger frequencies the self-energy due to interaction with a continuum gradually approaches the normal state result, obtained by summing up  series of logarithmical contributions from backscattering.

 Superconductivity at the onset of a nematic order occurs in the much studied FeSe$_{1-x}$S$_x$
   -- nematic order at $x \leq
    20\%$ (see, e.g., \cite{B_hmer_2017} and Refs.\ within). Our results for the spectral function can be verified in ARPES measurements around the critical doping.

 \textbf{Model and normal state results}.--- We consider a system of 2D itinerant spin-1/2 fermions on a square lattice near a QCP towards a
 $d$-wave nematic order  that breaks the four-fold rotation symmetry. We assume that 
 the interaction between fermions is mediated by soft fluctuations of the nematic order parameter.
  An appropriate Euclidean action is~\cite{PhysRevLett.91.066402}
\begin{align}
\notag
&\mathcal{S} =   \sum_\sigma\int_{k} \bar\psi_\sigma(k) (-i\omega_m + \xi_\k)\psi_\sigma(k) \ +
\frac{g}{V}  \sum_{\sigma, \sigma^\prime} \int_{k,p,q}  D(q) \times
 \\ &
f(\k)f(\p) \bar\psi_\sigma(k + \frac{q}{2}) \bar\psi_{\sigma^\prime} (p - \frac{q}{2}) \psi_{\sigma^\prime}(p+\frac{q}{2}) \psi_\sigma(k-\frac{q}{2}) \ .  \label{mainaction}
\end{align}
Here, $f(\k)$ is a form factor with $d_{xy}$ or $d_{x^2 - y^2}$ symmetry, $g$ 
is the effective fermion-boson coupling,
and  \begin{align} \label{Dintro}
 D(q) = - \frac{1}{M^2 + \aq^2 + g\Pi(q)} \ ,
 \end{align}
 where $M$ measures the small deviation from the QCP and $\Pi (q)$ is the dynamical 
      polarization, induced by the fermions, which has to be computed within the effective action of Eq.\ (\ref{mainaction}). {$D(q)$ should be seen as an effective interaction which includes both renormalizations from  low-energy [via $\Pi(q)$] and high-energy fermions, compare \cite{doi:10.1080/0001873021000057123}. }
 Eq.\ \eqref{mainaction} can be derived from, e.g., a multiorbital tight-binding model, rotated to the band basis, similar to Ref.\ \cite{Graser_2009}. The same effective action emerges in an isotropic system near a $d$-wave Pomeranchuk instability. In the latter case, 
 $f_{\q}(\k) = \cos(2\theta_{\k\q})$ ($\theta_{\k\q} = \measuredangle(\k,\q)$), invariant under simultaneous rotation of $\k,\q$. The key physics is the same in both cases, and  to simplify the presentation, we focus on the isotropic case and later discuss the changes introduced by the lattice.

\begin{figure}`
\centering
\includegraphics[width=.8\columnwidth]{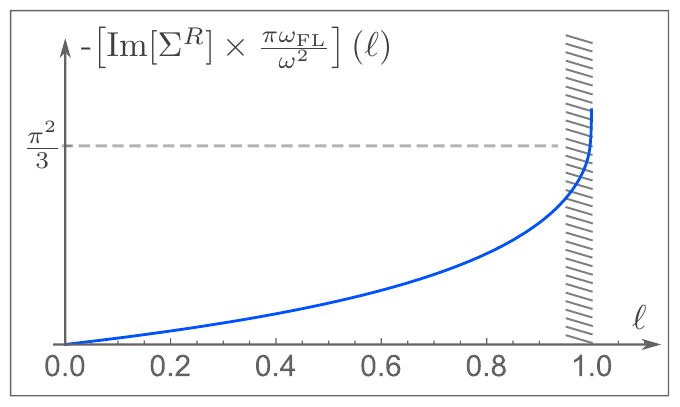}
\caption{Scattering rate divided by $\omega^2$ as given by Eq.\ \eqref{Sigmap2}, for $\epsilon = 0.05$.  Inside the dashed region, the normal state results are no longer valid. }
\label{fplot}
\end{figure}

The effective interaction $g D(q)$ leads to  much studied non-Fermi liquid behavior at $M \to 0$
     (Refs. \cite{PhysRevLett.91.066402,metlitski2010quantum,PhysRevB.80.165102,PhysRevB.81.045110})
        and
     also gives rise to an attraction in different pairing channels.
      The model  has two dimensionless parameters
   \begin{align}
\label{lambdadef}
\epsilon = \frac{M}{k_F} \quad   {\text{and}} \quad  \lambda = \frac{g}{4 \pi v_F M} \ .
\end{align}
The first one is a   measure of the distance to a QCP and the second {parametrizes the coupling strength}.
  In the immediate vicinity of a QCP, $\lambda$ is large and the system necessary falls into a strong coupling regime.  In our study we focus on the system behavior at small $\epsilon$, but in a  theoretically manageable Fermi liquid (FL)  regime $\lambda \ll 1$.
  To this end, we consider  the double limit $\epsilon \ll 1$ and $g \ll E_F$,  taken  such that $\epsilon \ll \lambda \ll 1$ {(which implies $\epsilon \gg g/E_F$, setting a lower bound on $M$ \footnote{{We note that the finite $M$ cuts off the logarithmic flow of $\lambda$, while right at the QCP this flow is cut off by the onset of pairing}}}). In this situation, the pairing instability develops at typical energies  of order $e^{-1/\lambda}$, which remain well inside the FL regime.
 {E.g., in FeSe$_{1-x}$S$_x$ the appropriate parameter window could be reached by increasing the sulfur doping $x$ or pressure beyond the quantum-critical value, where $\epsilon = 0, \lambda > 1$, which increases $M$ and therefore increases (decreases) $\epsilon$ ($\lambda$).} 
Our primary interest is the  retarded fermionic scattering rate Im[$\Sigma^R (k)]$ ($[G^R(k)]^{-1}  = [G^R_0(k)]^{-1}  + \Sigma^R(k)$, $k = (\omega,\k)$). We set $\k= \k_F$,   in which case
 $\Sigma^R(k) \equiv \Sigma^R  (\omega)$.
  We begin by listing the normal state results, which will serve as benchmarks
  for our calculations of Im[$\Sigma^R (\omega)$] in a superconductor.  In the normal state, the polarization bubble at small frequencies has a Landau-damping form, $\Pi(q) = 2\rho|\omega_m|/(v_F q)$, where $\rho = k_F/(2\pi v_F)$ is the density of states per spin.  At the one-loop level  the scattering rate has   the familiar 2D FL form \cite{PhysRevB.57.8873,PhysRevB.71.205112}
 \begin{align}
\label{sigma1resultmain}
 \text{Im} [\Sigma^{R,(1)}(\omega)] = \frac{2\lambda}{\pi}
\frac{\omega^2}{\omfl} \ln\left(\frac{\omfl}{|\omega|}\right) .
\end{align}
The logarithm comes from forward-scattering and backscattering processes, which play
a special role in 2D (Ref.~\cite{PhysRevB.71.205112}),
  and
   $\omfl = v_F M^3/g \rho$   sets
the upper edge of the FL behavior; our analysis is valid at $\omega < \omfl$. 
 At higher loop orders higher powers of the logarithm from multiple backscattering processes appear. They collectively change the scattering rate to \cite{pimenov2021onedimensional}
  \begin{align}
\label{Sigmap2}
&\Ims = \frac{\omega^2}{\pi \omfl} \notag\\
&\times \left[  \frac{\epsilon^2}{4(1- \ell)} - \frac{ (1-\ell)\ln^2(1-\ell)}{\ell} + 2\text{Li}_{2}(\ell)\right] \ ,
\end{align}
where $\ell = \lambda\ln(\omfl/|\omega|)$ and $\text{Li}_2 (\ell)$ is a polylogarithm. We plot $\Ims$ in Fig. \ref{fplot}.
  This expression becomes singular at $\ell = 1$, i.e.,  at
 $\omega \simeq \omfl e^{-1/\lambda}$
  and 
 is invalid at smaller $\omega$. The reason for this becomes clear once we note that the series of logarithms that transform (\ref{sigma1resultmain}) into (\ref{Sigmap2}) come from backscattering processes, which involve vanishingly small total momentum and are therefore the  same as  Cooper processes. The increase of  
 $\Ims/\omega^2$ with decreasing $\omega$ implies that the pairing interaction is attractive, hence the ground state must be a superconductor, and at  $\ell =1$, $\omega$ is of order of an expected pairing gap  $\Delta = \omfl e^{-1/\lambda}$.
 This clearly shows that the apparent singularity in Eq.\ (\ref{Sigmap2}) is the consequence of ignoring the fact that the ground state is a superconductor.
Below we start from a superconductor and show
 that the scattering rate is free of singularities at $\omega = O(\Delta)$.
\begin{figure}[b]
\centering
\includegraphics[width=\columnwidth]{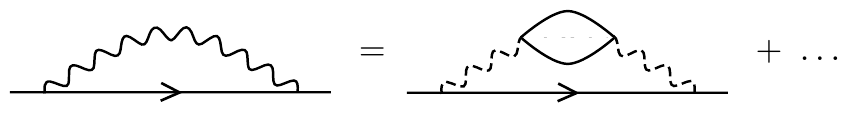}
\caption{Fermionic self-energy from excitation of a nematic fluctuation. Dashed wavy lines indicate a bare (static) nematic propagator.}
\label{basicdiag}
\end{figure}
To set the benchmark for the calculations in a superconductor, we make two observations. First, the full backscattering amplitude contains the sum of the contributions from different pairing channels,
 because the backscattering process
involves vanishingly small
total and transfer momentum.
 To  the
  leading order in $\epsilon$, partial amplitudes  in channels
 with different angular momenta are equal, and their sum  yields the two last terms in the r.h.s. of (\ref{Sigmap2}).  At $\ell =1$, this contribution tends to a constant $2\text{Li}_2 (1) = \pi^2/3$, although its derivative is  singular at $\ell =1$.  Beyond the leading order,  the $s$-wave component of the scattering amplitude is the largest, and the   $s$-wave pole in the backscattering amplitude gives rise to the first term in the r.h.s.\  of (\ref{Sigmap2}).  Second, Eq.\ (\ref{Sigmap2})  is obtained with the logarithmic accuracy and is valid up to $1-\ell = O(\lambda)$. At the edge of applicability, the $s$-wave pole contribution is of order $\epsilon^2/\lambda$ and is    small compared to the combined contribution from all     pairing channels, i.e.,       $\Ims  \simeq \frac{\Delta^2}{\omfl}$.  Below we compare this expression with the result that we obtain starting from a       superconducting  ground state.

 \textbf{Nematic propagator in the superconducting state}.--- The scattering rate in a superconductor is determined by the same process as in the normal state -- a decay into a nematic fluctuation and another fermion (Fig.~\ref{basicdiag}) \footnote{For sufficiently long-range interactions (small $M$), the rate may be influenced by additional collective low-energy excitations, see Refs.\ \cite{ KleinNpj2019, PhysRevB.62.11778}.}. However, in a superconductor both the fermionic propagator $G(q)$ and the polarization $\Pi(q)$ for  nematic fluctuations  get modified by the pairing gap $\Delta$. The modification of $G(q)$  is well-known \cite{Abrikosov2012}, but the modification of $\Pi (q)$ is model specific. In our case, a straightforward analysis yields
    \begin{align}
    \label{Piqmain}
& \Pi(q) = 2 \int_p f_\q(\p)^2 \times \\& \left[ G(p + q/2) G(p-q/2) - F(p+q/2) F(p-q/2) \right] \ , \notag
 \end{align}
  where
  $q = (\Omega,\q)$ and $G$ and $F$ are normal and anomalous Green's functions.
   The sign of the $FF$ term is the same as for the
   spin polarization operator at    ${\q} = (\pi,\pi)$ in a $d$-wave superconductor \footnote{In the latter case, there are two additional minus signs that compensate each other: one due to the presence of spin $\sigma$ matrices in the vertices, and another due to the sign change of a $d$-wave gap between $\k$ and $\k + \q$.}.
    Like there, the dynamical  bosonic propagator $D^R(\q=0)$
    then contains a resonance
    at $\Omega < 2 \Delta$,
   when
    $g\text{Re}[\Pi(\Omega, 0)] + M^2 = 0$ \cite{PhysRevLett.116.017001}.
   One can view the nematic resonance as a particle-hole bound state within the two-particle gap; it originates from a residual   interaction between Bogoliubov quasiparticles in a superconductor, which is attractive, like the original 4-fermion pairing interaction    \eqref{mainaction}.
     We note that such   resonance does not appear in the density-density correlator
     (a dressed particle-hole bubble with $s$-wave  form-factors), as there
     $\Pi(\Omega, \q = 0)$ has to vanish to respect charge conservation \cite{PhysRevB.94.064512,PhysRevB.96.014503}. On a technical level,  this holds because the $GG -FF$ contribution cancels out by the contribution  containing $GF$ combinations and vertex corrections~\cite{PhysRevB.96.014503}. For the nematic correlator that we consider, there is no cancellation and the physics can be captured by keeping only $GG-FF$ terms 
    (see the  supplementary material (SM) \cite{sm} for more discussion on this).  

\begin{figure}[b]
\centering
\includegraphics[width=\columnwidth]{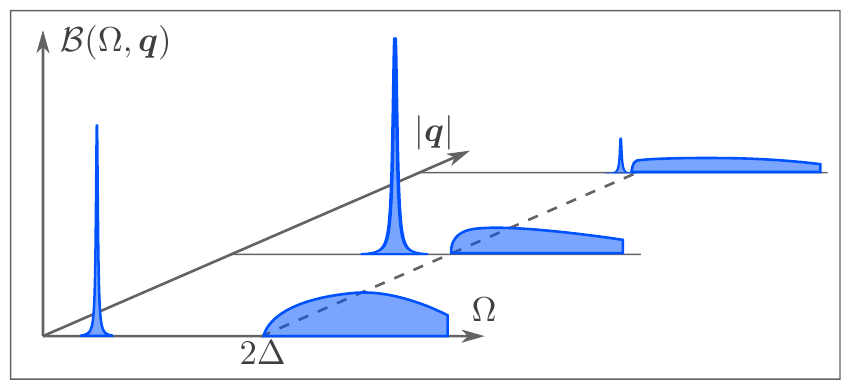}
\caption{Sketch of the nematic fluctuation spectrum $\mathcal{B}(\Omega,\q)
 = - {\text{Im}} [D^R (\Omega, \q)]$.}
\label{Bsketch}
\end{figure}

  We present the full analysis of $\Pi^R(q)$ in our case in the SM \cite{sm} and here quote the results.
   We have, at $\Omega \ll \Delta$,
   \begin{align}
\label{polePi1}
& \Pi^R (q) =\begin{dcases}&\hspace{-1em}-\frac{\rho}{6} \times  \frac{\Omega^2} {\Delta^2} - i0^+\sign(\Omega), \quad \ v_F\aq \ll \Delta  \\&\hspace{-1em}-\frac{\pi \rho}{4} \times \frac{ \Omega^2}{\Delta v_F \aq} - i0^+\sign(\Omega),  \quad  \ v_F\aq \gg \Delta  \end{dcases}  .
\end{align}
The existence of the two regimes is specific to our case of small-$\q$ fluctuations and originates from the $1/\aq$-dependence of the Landau damping in the normal state. Substituting $\Pi^R (q)$ into $D^R (q)$, we find that $\mathcal{B}= -\text{Im} [D^R]$ consists of a resonance (a pole) and a continuum, $\mathcal{B} (q) = \mathcal{B}_\text{res} (q) +  \mathcal{B}_{ \text{cont}}(q)$ -- see Fig.~\ref{Bsketch}.
 The resonance   describes
   a 2D plasmon-like propagating quasiparticle whose position  and the spectral weight strongly depend on the ratio $v_F \aq /\Delta$.  We find
\begin{align} \label{reseq}
&\mathcal{B}_{\text{res}}(q) = a_\q \times \delta(\Omega - \Omega_{\text{res}}(\q))  \\ &
\Omega_{\text{res}}(\q) \sim \begin{cases} \Delta \times  \sqrt{\epsilon/\lambda} &\ v_F \aq \ll \Delta  \\ \sqrt{\Delta v_F \aq} \times \sqrt{\epsilon/\lambda}  &\ v_F \aq \gg \Delta \end{cases}, \notag
\end{align}
and $a_\q$ varies between $a_\q \sim \Delta $ at $v_F \aq \ll \Delta$ and  $a_\q  \sim \sqrt{\Delta v_F \aq}$ at  $v_F\aq \gg \Delta$.  At small $\epsilon$ and $v_F \aq \sim \Delta$,  $\Omega_{\text{res}}(\q)$ is
 much smaller than $\Delta$.

The continuum $\mathcal{B}_{\text{cont}}(q)$ sets in at $\Omega = 2\Delta$ (see~\cite{sm} for details).
The full $\mathcal{B}(\Omega, \q)$ is sketched in Fig. \ref{Bsketch}. For $v_F \aq \sim \Delta$, the resonance is well separated from the continuum, but once
$v_F \aq$ exceeds
$\Delta  (\lambda/\epsilon)$,
 the resonance approaches  the continuum, and its  weight becomes exponentially small. {Likewise, a small-weight resonance close to $\Omega = 2\Delta$ is obtained for very weak coupling when $\epsilon/\lambda$ becomes too large.} \footnote{{For $\epsilon/\lambda$ > 1, the resonance condition can still be fulfilled below the continuum threshold due to divergence of $\Pi^R(\Omega \nearrow 2\Delta)$, see the SM \cite{sm}}}

\begin{figure}
\centering
\includegraphics[width=.95\columnwidth]{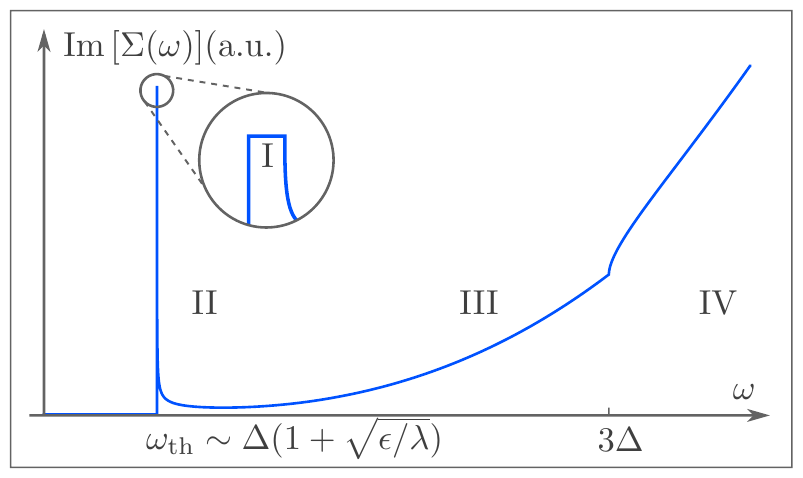}
\caption{Fermionic scattering rate, with scaling regimes as in Eq.~\eqref{mainformula}. Used parameters: $ \lambda = 0.4, \epsilon = 0.06$.}
 \label{ImSigmafig}
\end{figure}

 \textbf{Fermionic scattering rate at $\omega \sim \Delta$}. ---  With $\mathcal{B}(\Omega, \q)$   at hand,
   one can straightforwardly compute the scattering rate   using the spectral representation. We present the details in the SM and here list the results
   and   present the plot of
    $\text{Im}[\Sigma^R(\omega)]$,
    Fig.~\ref{ImSigmafig}.
    We obtain
 \begin{widetext}
     \begin{align}
     \label{mainformula}
&\text{Im}[\Sigma^R(\omega)]  \sim \begin{dcases}    \lambda \Delta \left(\frac{\epsilon}{\lambda}\right)^{{1/4}}\ ,  & 0 < \omega - \omt \lesssim    \frac{\Delta^3}{\omfl^2} \left(\frac{\epsilon}{\lambda}\right)^{5/2}  \hspace{1.83cm} \text{regime I}\\
  \frac{\Delta^2}{\omfl} \left(\frac{\epsilon^3}{\lambda}\right)^{1/2} \left(\frac{\Delta}{(\omega - \omt)}\right)^{1/2}\hspace{-1em},  &\frac{\Delta^3}{\omfl^2} \left(\frac{\epsilon}{\lambda}\right)^{5/2} \lesssim\omega - \omt \lesssim  \Delta \left(\frac{\epsilon}{\lambda}\right)^{1/2} \,   \quad \,  \text{regime II}\\
{ \frac{\lambda \Delta^2}{\omfl} }\left(\frac{(\omega - \omt)}{\Delta}\right)^{5/2}, & \Delta\left(\frac{\epsilon}{\lambda}\right)^{1/2} + \omt \lesssim \omega  \lesssim \Delta \left(3 + O(\lambda^2)\right)  \quad \text{regime III }\\
    \frac{\Delta^2}{\omfl}   \left(\frac{\omega - 3\Delta}{\Delta}\right)^{1/2}, &\omega \gtrsim \ \Delta\left(3 + O(\lambda^2)\right) \hspace{3.13cm}  \text{regime IV}
     \end{dcases}
 \end{align}
\end{widetext}
The scattering rate described by Eq.\ \eqref{mainformula} starts a threshold  $\omt = \Omega_{\text{res}}(0) + \Delta$, where it jumps to a value that is larger than the benchmark normal state expression $\Ims \sim \Delta^2/\omfl$ by a large factor $\sim E_F/\Delta$. It remains almost constant in a narrow energy window (regime I) and decreases as $1/\sqrt{\omega - \omt}$ (regime II), mirroring the familiar DoS enhancement in a superconductor. For increasing energies, the rate crosses through a minimum and grows again (regime III), which is due to the momentum dependence of the resonance position and weight. Finally, for $\omega > 3\Delta$ (regime IV), the rate acquires an additional contribution from scattering into the continuum, which gives rise to a kink in $\Ims$. For $\omega = 3\Delta + O(\Delta)$, the rate {becomes of the same order as  the normal state result and gradually transforms} into $\Ims$ from  Eq.\ (\ref{Sigmap2}).

\begin{figure}
\centering
\includegraphics[width=\columnwidth]{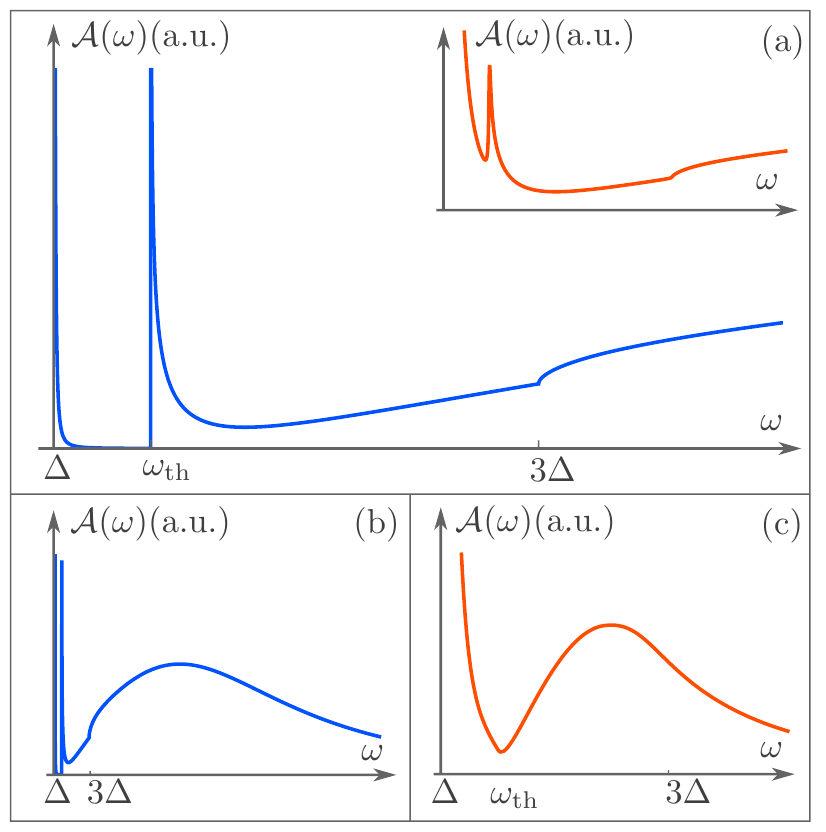}
\caption{Fermionic spectral function $\mathcal{A}(\omega)$ with same parameters as in Fig.\ \ref{ImSigmafig}. (a) $\mathcal{A}(\omega)$ for
 $\omega \lesssim 3\Delta$, showing the ``peak-peak-dip'' structure. In the inset, $\mathcal{A}$ is broadened with a phenomenological quasiparticle width $\Gamma = 0.2 \lambda \exp(-1/\lambda)$, due to additional processes such as impurity scattering. (b) $\mathcal{A}(\omega)$  at larger frequencies, showing the ``hump''. (c) \textit{Sketch} of $\mathcal{A}$ for $\omega \lesssim 3\Delta$, $\Gamma > 0$ and strong coupling $\lambda > 1$, showing conventional ``peak-dip-hump'' features.}
\label{typspectra}
\end{figure}

In Fig.\ \ref{typspectra} we plot the fermionic spectral function $\mathcal{A}(\omega) = (1/\pi)\text{Im}[G^R (\omega)]$ at  $\omega > \Delta$.  At $\lambda \ll 1$, $\mathcal{A}(\omega) \propto \Ims$.
   In this situation the most prominent   feature,   besides a sharp quasiparticle peak
 at  $\omega =\Delta$,
   is a narrow peak at $\omega_{\text{th}}$, followed by a broad minimum and  a cusp  at $3\Delta$.
 At larger frequencies where the normal state results are valid, a ``hump'' develops. This behavior can be termed as ``peak-peak-dip-hump". At $\lambda \geq 1$,  where $\Ims > \Delta$ at $\omega \geq \omt$, we      expect that the peak at $\omega_{\text{th}}$ is suppressed, as in this situation $A(\omega) \propto 1/\Ims$. Then the spectral function  displays a ``peak-dip-hump'' behavior,  similar to the one in the cuprates \cite{doi:10.1080/00018730600645636}.
 Since $\lambda \propto 1/M$, we expect a crossover from the peak-peak-dip-hump  lineshape to the peak-dip-hump one as the system comes closer to a nematic QCP.

 We now analyze how the results are modified for lattice systems. For $d_{x^2-y^2}$ symmetry, the nematic  form factor is  $f(\k) = \cos{k_x} - \cos{k_y}$; it reduces to $f(\k) \propto \cos(2\theta_{\k {\bf n}_x})$ for small $k_F$.
Such a form factor creates hot and cold regions on the Fermi surface. Our results for $\Sigma (\omega)$ are valid for the hot regions. The results hold with two modifications: (i) the jump of $\text{Im}[\Sigma (\omega)]$ at $\omega_{\text{th}}$ is reduced by the relative width of the hot region and (ii) there
exists a small but finite background $\text{Im}[\Sigma (\omega)]$ from scattering into cold regions, where the gap is small \cite{PhysRevB.98.220501}.

\textbf{Conclusion}.---
We analyzed the propagator of soft nematic fluctuations and 
 fermionic  scattering rate $\Ims$ of fermions in an
 $s-$wave SC state  near a nematic QCP.  We argued that
  $\Ims$   
   is qualitatively different from that
    in an ordinary $s$-wave SC  because 
      near a QCP, the pairing interaction in the $s$-wave channel is only slightly larger than in pairing channels with other angular momenta. We showed that the spectrum of nematic fluctuations in a SC consists of a  gapped continuum and  a strongly dispersing resonance mode.   We found three different regimes of the scattering rate, due to the interaction with a resonance: a jump at the threshold, a decrease towards a shallow minimum, and a slow subsequent increase.
  A fourth regime is realized at $\omega \geq 3\Delta$, when the scattering rate predominantly comes from the interaction with the continuum.
 At higher frequencies, $\Ims$ gradually transforms  into the normal state expression.  This leads to a ``peak-peak-dip-hump'' structure of the fermionic spectral function. This  structure transforms into the more familiar peak-dip-hump
 lineshape as the system comes closer to a nematic QCP and the coupling constant $\lambda$ increases.

 Superconductivity at the onset of a nematic order at $T=0$ develops in several systems studied in recent years, most notably in FeSe$_{1-x}$S$_x$ around $x=0.2$.  We call for ARPES measurements of the fermionic  spectral function to verify our results.

 \textbf{Acknowledgment}.---The work by A.V.C.\ was supported by the Office of Basic Energy Sciences, U.S. Department of Energy, under award  DE-SC0014402. A.K. was supported by NSF grant DMR-2037654.

 \clearpage  \newpage

\section*{Supplementary Material for ``Spectral signatures of a nematic QCP in high-$T_c$-superconductors''}

In this supplement, we present the detailed derivation of our results. In Sec.\ \ref{polfuncsec} we compute the main contributions to the polarization function $\Pi$. In Sec.\ \ref{GFsec}, we discuss the $GF$-contributions to $\Pi$. In Sec.\ \ref{nemflucsec}, we evaluate the nematic fluctuation spectrum $\mathcal{B}$. The self-energy $\Ims$ is derived in Sec.\ \ref{imsigmasec}. In Sec.\ \ref{SDWcompsec}, we compare our computation to the SDW case.

\section{Polarization function $\Pi$}
\label{polfuncsec}

\begin{figure}
\centering
\includegraphics[width=.95\columnwidth]{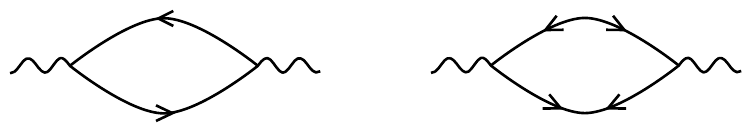}
\caption{Bubble diagrams contributing to the polarization function $\Pi$. }
\label{Pibubbles}
\end{figure}

Consider the polarization function $\Pi$, given by the two diagrams of Fig.\ \ref{Pibubbles}; the right-hand diagram represents the contribution from the anomalous propagators. $\Pi$ is given by:
\begin{align}
& \Pi(q) =   2 \int_p  f_{\q}^2(\p) \times \\ &  \left[ G(p + q/2) G(p-q/2) - F(p+q/2) F(p-q/2) \right] \ , \notag
\end{align}
where $q = (\Omega_m, \q)$ and the normal and the anomalous Matsubara propagators are defined as
\begin{align}
\label{Gmeanfieldmain}
G(p) = G(\tilde\omega_m, \p) &= -
 \frac{i \tilde\omega_m + \xi_\p}{\tilde\omega^2_m + \xi_\p^2 + \Delta^2}\ ,  \\ \notag    F(p) = F(\tilde\omega_m, \p) &=  \frac{\Delta}{\tilde\omega^2_m + \xi_\p^2 + \Delta^2} \ .
  \end{align}

We first evaluate $\Pi$ on the Matsubara axis  and then analytically continue to the real axis.
We compute the momentum integrals first --  in this way we omit a constant term $\lim_{\q \rightarrow 0} \lim_{\Omega_m \rightarrow 0}  \Pi(\Omega_m,\q)$, which renormalizes the  bosonic mass $M$ only.
 We find
\begin{align}
\label{PiAform}
&\Pi(q) = 2\rho\! \int \frac{d\theta_{\p\q}}{2\pi} f_\q(\p)^2 \int_{0}^\infty d\tilde{\omega}_m  \\ \notag &
\frac{(A_{+} + A_{-}) \times (A_{+} A_{-} + (\Omega_m)^2/4 - (\tilde\omega_m)^2 - \Delta^2)}{ A_{+} A_{-} \left(\left(A_{+} + A_{-}\right)^2 + (v_F \aq \cos(\theta_{\p\q}))^2\right)} =  \\&\notag  2 \rho\! \int_0^\infty d\tilde\omega_m \frac{A_{+} A_{-} + (\Omega_m)^2/4 - (\tilde\omega_m)^2 - \Delta^2}{ A_{+} A_{-} \left[ \left(\left(A_{+} + A_{-}\right)^2 + (v_F \aq)^2\right)^{1/2}+ R\right]} \ , \\ &A_\pm = \sqrt{ (\tilde{\omega}_m \pm \Omega_m/2)^2 + \Delta^2} \  . \notag
\\ &R = A_- + A_+ + \frac{(A_{-} + A_{+})  \times (v_F \q)^2}{2(A_+ + A_{-})^2 + (v_F \q )^2} \ ,  \notag
\end{align}
The last term $R$ originates from the nematic form factor $f_\q(\p) = \cos(2\theta_{\p\q})$. To obtain analytical results, we will focus on the limiting cases $v_F \aq \gg \Omega_m, \Delta$ or $v_F \aq \ll \Omega_m, \Delta$.
 In the first case, we can pull out $1/(v_F \aq)$ from the integral, and $R$ is not important: the angular integral in Eq.\ \eqref{PiAform} is dominated by $\theta_{\p\q} = \pm \pi/2$, and $f_\q(\p)^2 \simeq 1$. In the second case, we can neglect the $v_F \q$-terms, and inclusion of $R$ yields a prefactor $1/2$: the angular integral comes from all angles $\theta_{\p\q}$. We will take these factors into account implicitly and drop $R$ in the following.

In the normal state where $\Delta = 0$, for $v_F \q \gg \Omega_m$ we find, as expected
\begin{align}
\label{Pins}
\Pi(q)\big|_{\Delta = 0} = 2 \rho \frac{|\Omega_m|}{v_F \aq} \ .
\end{align}
Analytical continuation $i\Omega_m \rightarrow \Omega$ yields
\begin{align}
\label{PiRnormalstate}
\Pi^R(\Omega, \q)\big|_{\Delta = 0} = -i2 \rho \frac{\Omega}{v_F \aq} \ .
\end{align}

Let us focus on the superconducting state, $\Delta >0$, and first let $|\Omega_m| \ll \Delta$, which determines the nematic resonance. We can rewrite the numerator of Eq.\ \eqref{PiAform} as
\begin{align} \notag
&A_{+} A_{-} + (\Omega_m)^2/4 - (\tilde\omega_m)^2 - \Delta^2 = \\& \frac{\Omega_m^2}{2} \left( 1 - \frac{4(\tilde\omega_m)^2}{(A_+ + A_-)^2} \right) \ .
\end{align}
Inserting this form into Eq.\ \eqref{PiAform}, we can extract the leading $\Omega_m$-dependence in the two relevant limits $v_F \aq \ll \Delta, v_F \aq \gg \Delta$. We obtain the retarded expressions
\begin{align}
\label{polePi1}
&\Omega \ll \Delta, \  v_F\aq \ll \Delta:   \\ &\notag  \Pi^R (\Omega,\q) =- \frac{\rho}{6} \times  \frac{\Omega^2} {\Delta^2} - i0^+\sign(\Omega), & \\ \notag
 \\ \label{polePi2} &\Omega \ll \Delta, \  v_F\aq \gg \Delta:  \\ & \notag  \Pi(\Omega,\q) =-\frac{\pi \rho}{4} \times \frac{ \Omega^2}{\Delta v_F \aq} - i0^+\sign(\Omega)\ .  &
\end{align}

Next, we look at intermediate real frequencies $\Omega \simeq \Delta$ ($\text{Re}[\Pi^R(\Omega)]$ is even in $\Omega$, $\text{Im}[\Pi^R(\Omega)]$ is odd). The most important threshold is $\Omega = 2\Delta$, where excitation of the particle-hole continuum becomes possible. The general strategy to obtain $\Pi^R(\Omega \gtrsim 2\Delta)$ is to perform the analytical continuation in Eq.~\eqref{PiAform} \textit{before} integration; this correctly yields the singular part of $\text{Re}[\Pi^R]$, and the imaginary part can be restored using Kramers-Kronig relations. To extract the leading  contribution, we can let $\Omega \rightarrow 2\Delta, \tilde\omega_m \rightarrow 0$ in all regular parts of the integral. Rescaling $x = \omega_m/\Delta$, we find
\begin{align}
\label{Pixformula}
&\text{Re}[\Pi^R(\Omega, \q)] =  - 2\sqrt{2}\rho  \times \\ & \notag  \int dx \frac{1}{\sqrt{\alpha^2 + 4x^2}} \frac{1}{\left(\sqrt{\alpha^2 + 4x^2} + (x^2 - \alpha)+ \frac{(v_F \aq)^2}{2\Delta^2} \right)^{1/2}} ,
\end{align}
where
\begin{align}
\alpha = (\Omega - 2\Delta)/\Delta \ .
\end{align}
Consider first the limit $v_F \aq \gg \Delta, \Omega$. Then, for $\alpha \ll 1$:
\begin{align}
\text{Re}[\Pi^R(\Omega, \q)] = -\frac{2\rho  \Delta}{v_F \aq} \ln\left(\frac{v_F \aq}{(|\Omega - 2\Delta|) } \right),
\end{align}
and the corresponding imaginary part reads
\begin{align}
\text{Im}[\Pi^R(\Omega, \q)] = -\frac{2\pi \rho  \Delta}{v_F \aq} \times \theta(\Omega - 2\Delta) \ .
\end{align}
In the limit $v_F \aq \ll \Delta$, the continuum has additional structure, but this is of somewhat academic interest. For completeness, we give an order-of-magnitude estimate as obtained from Eq.\ \eqref{Pixformula}:
\begin{align}
&\alpha \ll (v_F \aq)^2/\Delta^2:
\\ \notag & \Pi^R(\Omega, \q) =  \\ \notag & -\frac{2\rho  \Delta}{v_F \aq} \ln\left(\frac{v_F \aq}{(|\Omega - 2\Delta|) } \right)  -i\frac{2\pi \rho  \Delta}{v_F \aq} \times \theta(\Omega - 2\Delta)   \\ \notag   \\ &(v_F \aq)^2/\Delta^2\ll  \alpha \ll 1 \label{sqrteq}:  \\  \notag & \Pi^R(\Omega, \q) \sim - \rho  \sqrt{\frac{\Delta}{\Omega - 2\Delta}}  \ln\left(\frac{\Delta}{v_F \aq} \right) (1+ i) \ .
\end{align}
The $1/(\Omega - 2\Delta)$ divergence in Eq.\ \eqref{sqrteq} agrees with the $(\q = 0)$-evaluation of $\Pi$ \cite{PhysRevLett.116.017001}. At the largest frequencies $\Omega \gg \Delta$, we recover the normal state result.

\begin{figure}
\centering
\includegraphics[width=\columnwidth]{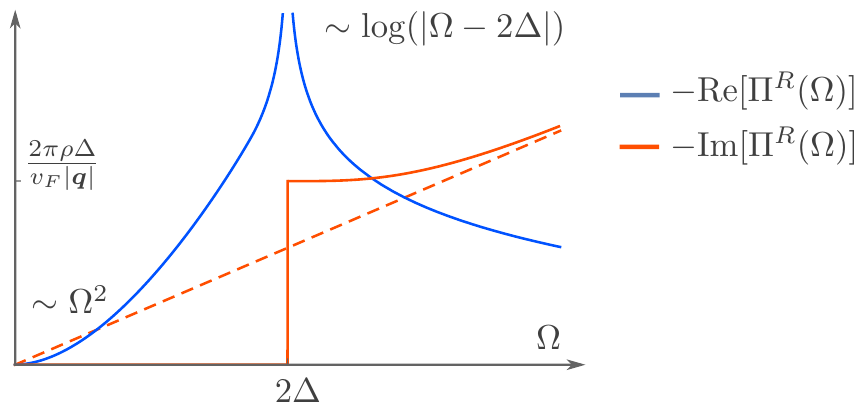}
\caption{Sketch of $\Pi (\Omega, v_F \aq \gg \Delta)$. The dashed line indicates the normal state result for $-\text{Im}[\Pi^R(\Omega,v_F \aq \gg \Delta)]$.}
\label{Piphsketch}
\end{figure}

A sketch of $\Pi^R(\Omega, v_F \aq \gg \Delta)$ is shown in Fig.\ \ref{Piphsketch}; $\Pi^R(\Omega, v_F \aq \ll \Delta)$ looks similar. The form of $\Pi^R$ is also similar to the SDW case \cite{ABANOV2001129}, but the crucial difference is  the overall scaling with $\Delta/ (v_F \q)$ which is absent in the SDW polarization function.

\section{$GF$-contributions}
\label{GFsec}

\begin{figure}[t]
\centering
\includegraphics[width=.6\columnwidth]{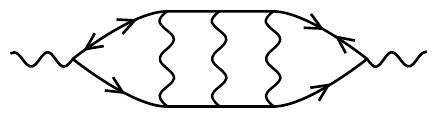}
\caption{Polarization diagram containing a renormalized particle-particle vertex. }
\label{GFdiag}
\end{figure}

In principle, the polarization $\Pi$ also receives contributions from diagrams as shown in Fig.\ \ref{GFdiag}, which contain a combination of $GF$-propagators and a renormalized particle-particle vertex. A priori, these contributions are not small. Without the $d$-wave form factor,  for external momentum $\q =0$ the diagrams of this type cancel the ($GG - FF$) contribution derived in the previous section, as dictated by charge conservation \cite{PhysRevB.94.064512,PhysRevB.96.014503}.

When the form-factor is included,  the diagram of Fig.\ \ref{GFdiag} vanishes
 for momentum-independent ($s$-wave) pairing interaction,
as
 it contains integrations over a single form factor $\propto \int d\phi \cos(2\phi) = 0$. In our case
 the pairing interaction  also contains a $d$-wave component, for which the integral does not vanish. 
  Evaluating the corresponding ladder diagram, we find
  \begin{equation}
\Pi_{GF}(\Omega, \q = 0) \sim - \rho \frac{\Omega^2}{\Delta^2} \frac{\lambda}{2\epsilon - \lambda \Omega^2/\Delta^2} \ .
\label{GFapp}
\end{equation}
We do not consider this contribution further for two reasons: (i) it has the same sign as the $(GG - FF)$-contribution, Eq.\ \eqref{polePi1}, in the low-energy limit, thus it can only enhance (and not destroy) the nematic resonance.
 (ii) the overall factor in (\ref{GFapp}) must be small, otherwise the system at $T=0$ would develop a $d$-wave pairing in addition to $s$-wave.    
  Such $s/d$-wave admixture without nematicity is a fairly exotic effect and will likely be suppressed once we include into consideration a repulsion between $s$- and $d$-wave order parameters.

\section{Nematic fluctuation spectrum $\mathcal{B}$}
\label{nemflucsec}

With $\Pi$ at hand, we can determine the nematic fluctuation spectrum as
\begin{align}
\mathcal{B}(\Omega,\q) = - \text{Im} \left[ - \frac{1}{M^2 + \aq^2 + g \Pi^R(\Omega, \q) } \right]  \ .
\end{align}
Inserting expressions \eqref{polePi1}, \eqref{polePi2}, we obtain the pole-like resonance part of the spectrum as
\begin{align}
\label{Bresdef}
&\mathcal{B}_{\text{res}}(\Omega, \q) = a_\q \times \delta ( \Omega - \Omega_{\text{res}}(\q)) \\
&\Omega_{\text{res}}(\q) \simeq  \begin{cases} \Delta \times  \sqrt{6} \kappa \left(1 + \frac{\aq^2}{M^2} \right)   &\ v_F \aq \ll \Delta  \\ \sqrt{\Delta v_F \aq} \times \sqrt{\frac{8}{\pi}} \kappa \left(1 + \frac{\aq^2}{M^2} \right)  &\ v_F \aq \gg \Delta   \end{cases}  \\ &
a_\q  \simeq  \begin{cases}  \pi \sqrt{\frac{3}{2}}  \times \frac{ \Delta}{2\kappa g \rho }    &\ v_F \aq \ll \Delta  \\ \sqrt{2\pi} \frac{\sqrt{\Delta v_F \aq}}{2\kappa g \rho }   &\ v_F \aq \gg \Delta  \ .  \end{cases}
\end{align}
Here, we have introduced
\begin{align}
\kappa = \sqrt{\frac{M^2}{2g \rho }} = \sqrt{\frac{\omfl}{v_F M }} = \sqrt{\frac{\epsilon}{4 \lambda}} \ll 1\ ,
\end{align}
and also assumed $\aq \ll M$. Note that the weight of the resonance peak increases with $v_F \aq$.
The above expressions are valid as long as $\Omega_{\text{res}} < 2\Delta$, which constraints the maximal allowed $\aq$-values to $v_F \aq < \Delta/\kappa^2$. For $\Omega > 2\Delta$, the continuum part $\mathcal{B}_{\text{cont}}$ sets in. While we do not discuss the form of the continuum in detail, notice the following trends that are reflected in Fig.\ \ref{Bsketch} of the main text:

(i) As long as the resonance is well-separated from the continuum, always $ \Pi^R(2\Delta + 0^+) \gg M^2 $, thus $\mathcal{B}_\text{cont}$ does not jump at $\Omega = 2\Delta$, but sets in continuously. However, as the resonance approaches the continuum, $\Pi^R(2\Delta + 0^+) > M^2$ only due to the logarithmic enhancement of $\text{Re}[ \Pi^R]$. Eventually, for $v_F \aq \gg \Delta/ \kappa^2$, $M^2$ becomes the dominant scale, which means that $\mathcal{B}_\text{cont}$ jumps at $\Omega = 2\Delta$. At these large momenta, we can approximate
\begin{align}
\label{Bcontsupp}
\mathcal{B}_{\text{cont}}(\Omega, \q) \simeq  \frac{2\pi g \rho }{M^4} \times \frac{\Delta}{v_F \aq}  \times \theta(\Omega - 2\Delta) \ .
\end{align}

(ii) For small momenta, the maximum value of the continuum $\mathcal{B}_\text{cont} \simeq 1/M^2$ is expected at frequencies $\Omega - 2\Delta = O(\Delta)$ when $\Pi^R \simeq M^2$. The weight of the continuum can be estimated as $\Delta/M^2$, which is larger than the weight of the resonance by about a factor of $\kappa$. As $\aq$ is increased, the weight goes down, as is also seen from Eq.\ \eqref{Bcontsupp}.

\section{$\text{Im}[\Sigma^R(\omega, \k)]$ for $\omega \lesssim \Delta$}
\label{imsigmasec}

We need to compute $\text{Im}[ \Sigma^R(\omega)] \equiv \text{Im}[ \Sigma^R(\omega, \k)]$, with $\k$ on the Fermi surface.
From the spectral representation, one can derive the following general formula
\begin{align}
\label{spectralrep}
&\text{Im}[\Sigma^R(\omega)] =  \\ & \notag g \!\int\! \frac{d\q}{(2\pi)^2} f^2_{\q}(\k + \q/2) \int_0^\omega \frac{d\Omega}{\pi} \mathcal{A}_0 (\omega - \Omega, \k +\q) \mathcal{B}(\Omega, \q)  \ .
\end{align}
We focus on $\omega > 0$.  $\mathcal{A}_0$ is the ``bare'' spectral function,
\begin{align} \notag
&\mathcal{A}_0(\omega - \Omega, \k + \q)  =-\text{Im}\left[G^R(\omega - \Omega, \k + \q) \right] =  \\ & \pi u_{\k+\q}^2 \delta\left( \omega - \Omega - E_{\k+\q}
\right),   \\ &  E_{\k + \q} = \sqrt{\Delta^2 + \xi_{\k + \q}^2}, \ u_{\k + \q}^2 = \frac{1}{2} \left( 1+ \frac{\xi_{\k +\q}}{E_{\k + \q}} \right) . \notag &
\end{align}
In Eq.\ \eqref{spectralrep}, we can approximate $f^2_\q(\k + \q/2) \simeq f^2_\q(\k) \simeq (-1)^2 = 1$, since the integral will be dominated by angles such that $\k \perp \q$.

We start with the contribution of the nematic resonance, inserting $\mathcal{B}_{\text{res}}(\Omega, \q)$ into Eq.\ \eqref{spectralrep}. Then
\begin{align}
&\text{Im}[\Sigma^R_\text{res}(\omega)]= \\ & \notag g\!\int\! \frac{d\q}{(2\pi)^2} u_{\k + \q}^2 a_\q \delta(\omega - \Omega_{\text{res}}(\q) - E_{\k + \q}) \ .
\end{align}
The resonance contribution is only important for $\omega < 3\Delta$, which implies that $\xi_{\k + \q}$ is at most of order $\Delta$ or smaller. To get the correct order of magnitude, we can therefore approximate $u_{\k + \q}^2 \simeq 1/2$ and expand $E_{\k + \q}$ in $\xi_{\k + \q} = v_F \aq \cos(\phi)$, where $\phi = \measuredangle(\k,\q)$. $\text{Im}[\Sigma^R_\text{res}]$ starts at the threshold $\omega = \omega_{\text{th}}$, where
\begin{align}
\omega_{\text{th}} = \Delta + \Omega_{\text{res}}(0) = \Delta(1+\sqrt{6} \kappa) \ .
\end{align}

 Close to $\omega_{\text{th}}$, we need the low-momentum branch of $\mathcal{B}_{\text{res}}$, cf.\ Eq.\ \eqref{Bresdef}. We get
\begin{align}
\label{Imsigmalowmom1}
&\text{Im}[\Sigma^R_\text{res}(\omega)] = \frac{\sqrt{6}}{16}  \frac{\Delta}{\kappa \rho} \int_0^{\Delta/v_F} d\aq \aq \int \frac{d\phi}{2\pi} \\ \notag & \delta (\omega - \omt - (v_F \aq \cos(\phi))^2/2\Delta - \sqrt{6}\kappa\Delta  \aq^2/2M^2) \ .
\end{align}
We have cut off the $\q$-integral at $\Delta/v_F$, which marks the upper boundary of the low-momentum branch of $\mathcal{B}_{\text{res}}$. The trivial $\aq$-integral yields
\begin{align}
\label{Imstothetas}
&\text{Im}[\Sigma^R_\text{res}(\omega)]  = \\ \notag  &\frac{\sqrt{6}}{16}  \frac{\Delta^2}{\kappa \rho  v_F^2} \times \int_0^{2\pi} \frac{d\phi}{2\pi} \frac{1}{\cos^2(\phi) + \sqrt{6} \kappa \Delta^2/ (v_F M)^2} \times \\ \notag & \theta\!\left( \Delta - \frac{\omega - \omt}{\cos^2(\phi) +  \sqrt{6} \kappa \Delta^2/ (v_F M)^2 } \right) \times \theta (\omega - \omt) \ .
\end{align}
First consider the limit $\omega \searrow \omt$, then the first step function in Eq.\ \eqref{Imstothetas} equals  one. The integral comes from angles close to $\pi/2, 3\pi/2$: the constant $\sqrt{6} \Delta^2/(v_F M)^2$ is extremely small, since $\Delta = \omfl \exp(-1/\lambda)$ is exponentially suppressed compared to $v_F M = \omfl/\kappa^2 $. In this limit, we find
\begin{align}
&\text{Im}[\Sigma^R_\text{res}(\omt +  0^+)] = \\ & \notag  \frac{6^{3/4}}{16} \frac{ \Delta M}{\kappa^{3/2} \rho  v_F} =  \notag \left(\frac{27}{32}\right)^{\!1/4} \times  (\kappa^{1/2}  \lambda) \times \Delta\ .
\end{align}
Moving away from the threshold, for $\omega - \omt \gtrsim \Delta \sqrt{6} \kappa \Delta^2/ (v_F M)^2$, the step-function becomes important: it restricts $\cos^2(\phi) \gtrsim (\omega - \omt)/\Delta$. Since the upper cutoff $\Delta$ enters explicitly, we no longer have control over numerical prefactors and  suppress them from now on. We obtain
\begin{align}
\label{index1res}
&\text{Im}[\Sigma^R_{\text{res},1}(\nu)] \sim \kappa^3 \frac{\lambda \Delta^2}{\omfl} \sqrt{\frac{\Delta}{(\omega - \omt)}} \ , \end{align}
where the index $1$ means that this is the low-$\q$ contribution. Now we compare this to the large-$\q$ resonance contribution of Eq.\ \eqref{Bresdef}. It reads
\begin{align}
&\text{Im}[\Sigma^R_{\text{res},2}(\omega)] \sim \\ & \int d\aq  d\phi \aq \frac{\sqrt{\Delta v_F \aq}}{\kappa  \rho} \delta( \omega - \kappa \sqrt{ \Delta v_F \aq} - E_{\k + \q} ) \  . \notag
\end{align}
To find the low-energy result, we require that $v_F \aq \gg \Delta$, but still $v_F \aq \cos(\phi) \ll \Delta$, s.t.\ we can expand $E_{\k + \q}$ as before. Rescaling momenta, $x = v_F \aq /\Delta$, and linearizing the angular integral around $\pi/2$ (variable $y$), we can write
\begin{align}
&\text{Im}[\Sigma^R_{\text{res},2}(\omega)] \sim \\ & \frac{\Delta^2}{\kappa \rho  v_F^2} \int dx \int dy \ x^{3/2} \times \delta\left(\frac{\omega - \Delta}{\Delta} - \kappa \sqrt{x} - x^2 y^2\right). \notag
\end{align}
The lower bound on the $x$-integral is $x_{\text{min}} = 1$. The delta-function sets
\begin{align}
|y| = \frac{\sqrt{\frac{\omega - \Delta}{\Delta} - \kappa \sqrt{x}}}{x} \ . \label{y0value}
\end{align}
To have $y x < 1 \  \forall x$ (i.e., $v_F \aq \cos(\phi) < \Delta$), we need $\omega - \Delta(1+\kappa) < \Delta$. Then, to have $|y| > 0$ in Eq.\ \eqref{y0value}, we must require $x < x_\text{max} = [(\omega - \Delta)/(\Delta \kappa)]^2$. The $y$ integral yields
\begin{align}
&\text{Im}[\Sigma^R_{\text{res},2}(\omega)] \sim \\ \notag  &  \frac{\Delta^2}{\kappa \rho  v_F^2} \int_1^{x_\text{max}} dx  \frac{\sqrt{x}}{\sqrt{\frac{\omega - \Delta}{\Delta} - \kappa \sqrt{x}}} =  \\ &   \frac{\Delta^2}{\kappa^{3/2} \rho  v_F^2} \int_1^{x_\text{max}} dx \frac{\sqrt{x}}{\sqrt{\sqrt{x_{\text{max}}} - \sqrt{x}}}  \ .
\notag \end{align}
This contribution becomes sizeable compared to $\text{Im}[\Sigma^R_{\text{res},1}]$ if $x_{\text{max}} \gg 1$ (but still $x_{\text{max}} < 1/\kappa^2$, otherwise $\omega - \Delta(1+\kappa) > \Delta$). We can then approximate $\omega - \Delta \simeq \omega - \omt$, and the integral yields
\begin{align}
\label{sigmares2final}
&\text{Im}[\Sigma^R_{\text{res},2}(\omega)] \sim  \frac{\Delta^2}{\kappa^{3/2} \rho  v_F^2} (x_{\text{max}})^{5/4} \\ &\sim \kappa^3 \frac{\lambda \Delta^2}{\omfl} \sqrt{\frac{\Delta}{\omega - \omt}} \times \kappa^{-3} \left( \frac{\omega - \omt}{\Delta}\right)^3 \ . \notag
\end{align}
Comparing to Eq.\ \eqref{index1res}, we see that Eq.\ \eqref{sigmares2final} becomes dominant for $(\omega - \omt)>  \kappa\Delta$.

So far, we only considered the contribution from the nematic resonance, inserting $\mathcal{B}_{\text{res}}$
into Eq.\ \eqref{spectralrep}. The contribution from the continuum $\mathcal{B}_{\text{cont}}$ has been evaluated in Ref.\ \cite{pimenov2021onedimensional}, see App.\ G within. The result is, for $\omega - 3\Delta \ll \Delta$:
\begin{align}
\text{Im}[\Sigma^R_{\text{cont},2}(\omega)] \sim   \frac{\Delta^2}{\omfl}  \times \theta(\omega - 3\Delta) \sqrt{\frac{\omega - 3\Delta}{\Delta}} \ .
\end{align}
Thus, the continuum contribution becomes larger than the resonance part  \eqref{sigmares2final} for $\omega - 3\Delta \gtrsim \lambda^2 \Delta$. Alltogether, the self-energy at energies $\omega \lesssim 3\Delta$ reads
\begin{align}
\label{sigmacasesapp}
&\text{Im}[\Sigma^R(\omega)]  \\ & \notag \sim \begin{dcases}
 \kappa^2 \epsilon \frac{\Delta^2}{\omfl}, & \Delta < \omega  < \omt \\
  \kappa^{1/2}  \lambda \Delta\ ,  & 0 < \omega - \omt \lesssim    \frac{\kappa^5\Delta^3}{\omfl^2}  \\
 \kappa^3 \frac{\lambda \Delta^2}{\omfl} \left(\frac{\Delta}{(\omega - \omt)}\right)^{1/2}\hspace{-1em},  &\frac{\kappa^5\Delta^3}{\omfl^2} \lesssim\omega - \omt \lesssim \kappa \Delta \\
 \frac{\lambda \Delta^2}{\omfl} \left(\frac{(\omega - \omt)}{\Delta}\right)^{5/2}, & \kappa \Delta + \omt \lesssim \omega  \lesssim 3\Delta + \lambda^2 \Delta \\
    \frac{\Delta^2}{\omfl}   \left(\frac{\omega - 3\Delta}{\Delta}\right)^{1/2}, &\omega \gtrsim \ 3\Delta + \lambda^2 \Delta
 \end{dcases}
 \end{align}
Here we have also added the minor Goldstone-mode contribution derived in Ref.\ \cite{pimenov2021onedimensional} at the smallest energies. Eq.\ \eqref{sigmacasesapp} is equivalent to Eq.\ (\ref{mainformula}) of the main text. 

\section{Comparison to the SDW case}
\label{SDWcompsec}

There are two qualitative differences between the self-energies due to excitation of the nematic resonance (Ising nematic case) and spin-resonances (SDW case): (i) in the SDW case, the superconducting ground state has $d$-wave order, and gapless states exist at the nodes along the Brillouin zone diagonal. Therefore,  $\Ims$ generically starts at $\omega = 0$. However, the nodal states cannot couple efficiently to the spin resonance due to a momentum mismatch (see, e.g., Ref.\ \cite{doi:10.1080/00018730600645636}), and therefore their contribution to the self-energy is small. (ii) the spin resonance is peaked at a large momentum $Q = (\pi,\pi) \sim k_F$ and not at $Q \simeq 0$. As a result, the quadratic part of the corresponding polarization function scales as \cite{ABANOV2001129}
\begin{align}
\Pi^{R}_{\text{SDW}}(\Omega) \sim \rho \frac{\Omega^2}{\Delta E_F} \ ,
\end{align}
while in the nematic case $\Pi^R \sim \rho \frac{\Omega^2}{\Delta^2}$. Thus, $\Pi^R$ is generically much larger in the nematic case. In the SDW case, one then finds that the spin resonance is located at
\begin{align}
\label{OmegaresSDW}
\Omega_{\text{res}}^{\text{SDW}} \sim \sqrt{\Delta \omega_{\text{SF}}} \ ,
\end{align}
where $\omega_{\text{SF}}$ is the analog of $\omfl$. For Eq.~\eqref{OmegaresSDW} to hold, one needs $\omega_{\text{SF}} < \Delta$. This shows that a generic spin resonance is only possible \textit{at strong coupling} $\lambda \gtrsim 1$, otherwise $\Delta \sim \omega_{\text{SF}} \exp(-1/\lambda) \ll  \omega_{\text{SF}}$.  Without the contribution from the nodal states, the SDW self-energy then starts at $\Delta + \Omega_{\text{res}}^{\text{SDW}}$. At larger energies, it becomes smaller, similar to the nematic case. However, with a calculation analogous to Eq.\ \eqref{Imsigmalowmom1} one can show that this reduction is only by a factor
\begin{align}
\sim \sqrt{\frac{\Delta \Omega_{\text{res}}^{\text{SDW}}}{(v_F M)^2}} = O(1) \ \quad \text{for} \quad \lambda = O(1) \ .
\end{align}
The imaginary part of the self-energy is therefore step-like and not a sharp peak; it looks similar to Fig. \ref{Piphsketch}. Thus, only the peak-dip-hump scenario can be realized in the SDW case, not the peak-peak-dip-hump scenario.

If the coupling $\lambda$ is small in the SDW case, a spin resonance is in principle still possible (see \cite{ABANOV2001129} and references within). This weak-coupling resonance occurs due to the logarithmic divergence of the polarization function at $\Omega = 2\Delta$, and appears at $\Omega_{\text{res}}^{\text{SDW}} = 2\Delta(1- Z_s)$, where
\begin{align}
Z_s = \exp (- \omega_{\text{SF}}/\Delta)
\end{align}
is also the weight of this resonance (similar to $a_\q$ in Eq.\ \eqref{Bresdef}). But for $\lambda \ll 1$, $\Delta/\omega_{\text{SF}}$ is exponentially small, thus $Z_s$ is ``superexponentially" small. An analogous small-weight resonance is indicated in Fig.\ \ref{Bsketch} of the main text at the largest momenta.

\bibliography{nematic_resonance}
\bibliographystyle{apsrev4-2}

\end{document}